# Scaling and allometry in the building geometries of Greater London†

Michael Batty[1], Rui Carvalho[2], Andy Hudson-Smith[1],
Richard Milton[1], Duncan Smith[1] and Philip Steadman[3]

[1]Centre for Advanced Spatial Analysis, University College London,
1-19 Torrington Place, London WC1E 6BT, UK

[2]School of Mathematical Sciences, Queen Mary, University of London, Mile End Road, London E1 4NS, UK

[3]Bartlett School of Architecture and Planning, University College London,
Wates House, Gordon Street, London WC1E 6BT, UK

m.batty@ucl.ac.uk, r.carvalho@qmul.ac.uk, asmith@geog.ucl.ac.uk,
richard.milton@ucl.ac.uk, duncan.a.smith@ucl.ac.uk, j.p.steadman@ucl.ac.uk

Abstract. Many aggregate distributions of urban activities such as city sizes reveal scaling but hardly any work exists on the properties of spatial distributions within individual cities, notwithstanding considerable knowledge about their fractal structure. We redress this here by examining scaling relationships in a world city using data on the geometric properties of individual buildings. We first summaries how power laws can be used to approximate the size distributions of buildings, in analogy to city-size distributions which have been widely studied as rank-size and lognormal distributions following Zipf [1] and Gibrat [2]. We then extend this analysis to allometric relationships between buildings in terms of their different geometric size properties. We present some preliminary analysis of building heights from the **Emporis** database which suggests very strong scaling in world cities. The data base for Greater London is then introduced from which we extract 3.6 million buildings whose scaling properties we explore. We examine key allometric relationships between these different properties illustrating how building shape changes according to size, and we extend this analysis to the classification of buildings according to land use types. We conclude with an analysis of two-point correlation functions of building geometries which supports our non-spatial analysis of scaling.

† *This research was supported by the UK Engineering and Physical Sciences Research Council (EPSRC under grant EP/C513703/1 and by the UK National Centre for e-Social Science in the GeoVUE Project (ESRC under grant RES-149-25-1023).*

## 1 Introduction

Cities are structured according to the rules of spatial competition which manifest themselves in self-similar patterns which are fractal. Populations tend to cluster around market locations which reflect a hierarchy of needs from the essential to the specialist, ordered spatially according to the strength of demand while densities tend to reflect economies of agglomeration which generate a small number of very high density locations and a large number of lower ones. The patterns that emerge are sustained by transportation routes that tend to fill space in the most economical way, minimizing length and maximizing capacity, whose spatial organization is usually hierarchical and tree-like. Cities are thus composed of fractal-like clusters on many spatial scales whose order appears to follow well-defined numerical rules of scaling.

Most demonstrations of such order in fact pertain to systems of cities rather the spatial organization of the city itself, focusing on size distributions in which spatial order is implicit [3]. The size distribution of cities in fact is scaling with Zipf's [1] rank-size rule acting as the bench-mark against which many other spatial distributions are compared and contrasted. Most of the work to date on city-size distributions throws away any spatial structure that exists. Cities measured by their populations, incomes or employment, are considered as dimensionless points with their sizes reflecting competition between whole cities rather than competition between their component parts. In essence, the fact that there are a small number of large cities and a large number of small and that this distribution manifests a regularity which appears persistent through time, reflects the consequence of competitive processes under resource limits: there is simply never the resources or demand to sustain large numbers of large cities, and thus most cities remain small. The same mechanisms clearly exist at the more local scale, within cities with the competition perhaps being less fierce but regular ordering of populations and other activities by size being the norm rather than the exception.

Inside cities, the predominant theory of ordering is based on a microeconomics that suggests that densities of population, rent, and employment decline with increasing transport costs from the most intensive hubs or clusters of economic activity [4]. It is easy to speculate that such order is consistent with a regular size distribution of population densities for hypothetical models where transport cost or distance from any point is equated with the rank order. But such research has never been followed up and we will simply note it in passing. Research on scaling distributions barely touches the spatial structure of the city where the focus has been much more on fractal patterns rather than their scaling structure [5]. In this paper, we will extend the study of size distributions to the internal structure of cities treating spatial structure only implicitly, demonstrating that scaling orders are as strong within cities as between, and then reintroducing space to show its relative importance.

There is an additional twist to our analysis of intra-city-size distributions for our focus here is on geometric rather than economic or demographic attributes of the city. We consider that scaling in cities is strongly related to the constraints that geometry imposes on density and nearness and thus we will examine the size distributions of buildings in terms of their Euclidean footprint – area, perimeter, height and so on – making the rather loose argument that these sizes reflect indirectly population and employment volumes. Moreover, as buildings grow in size, their shape must change to enable them to function and thus their scaling can be linked to their allometry. In fact a sound theory of urban allometry should relate social and economic activity to building geometry and in this paper, we hope to set the agenda for further work in this area.

To date, work on the scaling of activities in cities has been sparse. As remarked, the study of rank-size distributions across cities has been extensive and work on urban density profiles has been significant. But there has barely been any work on building geometries with the exception of Bon [6] and Steadman [7]. There has been some on transport and infrastructure supply networks [8-10] and some on the allometry of transport networks [11-13]. Currently, West, Brown, and Enquist [14] are beginning to apply their theory of metabolic scaling to cities and social systems [15-16], thus providing a marker for a better understanding of the way cities scale as they grow.

In the next section, we will introduce the idea of scaling as an approximation to some underling order in the size of things, relating this to ways of representing this order as densities and distributions, and we will link this to the key allometric relationships that characterize building geometry in terms of their volume, the area of their footprint, their height, and their perimeter. Our first foray into analysis looks at scaling in the height of buildings world-wide and in three cities – London, Tokyo and New York from the **Emporis** data base (http://www.skyscraper.com/). We show quite conclusively that these distributions can be well approximated by rank-size distributions that imply power laws. We then outline the main database that we are working with for Greater London which contains some 3.6 million building blocks. Analysis of this data then proceeds, first for rank-size scaling of building geometries, then for allometric relationships. We finally introduce two-point correlation measures of the spatial distribution of these building geometries demonstrating that the strong scaling relations already detected, are not completely destroyed when we extend the analysis to include their spatial extent.

## 2 Approximating urban order through rank-size scaling

It is over 100 years since distributions of objects and attributes characterizing human populations such as city sizes and incomes were first described using power laws [17, 18] with Zipf's [1] work popularizing the idea in the mid twentieth century as the rank-size rule. Since then, there has been a slow realization that a more likely form for such distributions is the lognormal with simple stochastic models, particularly those based on growth by proportionate effect due to Gibrat [2], finding favor as one of the generating mechanisms of such phenomena. The current conventional wisdom is that the power law is a good approximation to the distribution of the lognormal in its 'fat tail' which describes the form of the largest sizes in the distribution. We will follow this convention here, not seeking to fit building size distributions to the lognormal but assuming that they can be approximated as power laws. As we shall see, the distributions for Greater London in fact show little sign of lognormality and thus our assumption appears tenable. Although considerable effort has been made in fitting such distributions using the original Zipf rank-size relationships, these are directly related to the underlying density and cumulative densities of their size distributions [19, 20]. We will thus first introduce these transformations from their densities to their rank-size distributions providing a clear basis for our estimation procedures.

To illustrate the way we transform densities into distributions, we first define density $p_i$ where $i$ is the object in question, in this context, the location of a building. We order these locations from the smallest to the largest densities and thus change the index from $i$ to $k$. The density and distributions we work with are thus based on $p_k$ which follow the order from the smallest to the largest. To plot the probability density function (PDF), we usually bin the data, which for systems where the sizes of each object follow a power law, provide a distribution which is highly skewed to the left. In fact, there has been a long debate about whether such distributions follow a power law or a lognormal for many distributions resemble a highly skewed normal distribution where the power law is used to approximate the fat or heavy tail. Apart from noting this here [21, 22], we will not pursue it further as it is controversial when applied to the way buildings sizes are located, constructed and evolve in cities. The cumulative distribution function (CDF) defined as $F(p_K \le p_k)$ can be computed from the raw data as $\sum_{k=1}^{K} p_k$ without binning and it is thus preferable to work with the data in this form. This is equivalent to the integral of the continuous density. In fact, the normal practice in examining such size distributions is to use the counter or complementary-cumulative distribution function (CCDF) which we define here as $r = F(p_K \ge p_k) = N - F(p_K \le p_k)$. The CCDF is none other than the rank-size distribution defined by Zipf [1] and used extensively in approximating the fat tail as a power law. Note that henceforth $r$ will be used to define rank in terms of the ordered sizes defined by $k$.

To illustrate how we assume scaling in such distributions, we usually plot the CCDF on a logarithmic scale. This plot gives greater visual weight to the larger values of density and it is intuitively clear that the relationship can be approximated by a straight line which is the signature of a power law. We now approximate the power law as a continuous density suppressing the index $k$ as

$$f(p) \sim p^{-\alpha} \qquad (1)$$

where $\alpha$ is the power of the density. The cumulative (and counter-cumulative which has the same functional form) is the integral of (1) and is

$$F(p) \sim p^{-\alpha+1} \qquad (2)$$

which we can also write explicitly in rank-size terms as

$$r \sim p(r)^{-(\alpha-1)} \quad . \qquad (3)$$

The usual form however is where density is written as a function of rank. Then from equation (3)

$$p(r) \sim r^{-\frac{1}{\alpha-1}} = r^{-\beta} \quad , \qquad (4)$$

where we now define $\beta$ as the (inverse) power.

From equations (1) to (4), it is clear that if such an approximation is warranted, then the parameter of the density function $\alpha$ must be greater than 1 for the cumulative distribution function to be defined. If we logarithmically transform (4), we produce the linear equation

$$\log p(r) = \log G - \beta \log r \qquad (5)$$

which can be estimated in a straightforward manner using regression. In various applications, we have used the Hill maximum-likelihood estimator favored by Newman [20] although here we have kept to the traditional method of regression because as we will see, the scaling for the Greater London building geometries is so clear that we consider regression to be quite robust. We have not tested the degree to which these distributions are lognormal or scaling but Clauset, Shalizi, and Newman [23] have introduced a series of tests to enable this. In future work, we will follow this best practice but as this paper is simply an exploration of the extent to which scaling might be present in building geometries and allometry, we stick with current practice.

## 3 Allometry in urban size distributions

In previous work in measuring scaling in cities, the focus has been on populations and related attributes where individuals are aggregated into small zones or indeed entire cities whose size distribution shows scaling. However here our focus will be upon building sizes which do not need to be so aggregated and this tends to make the analysis somewhat more direct, hence simpler. We first need to define the geometric properties of buildings that we will use to measure their size. Consider a building to be an irregular block defined in terms of the appropriate lengths of its three dimensions. For each building, height $H_j$, the area of its footprint $A_j$, the perimeter of this area $L_j$, and the building volume $V_j = A_j H_j$ can be calculated directly although volume which is probably the best measure of size, is a product of all three dimensions, in turn a function of the area and height measures. Each of these has a rank order $r$ which we will test for scaling using power law approximations $H(r) \sim r^{-\beta_H}$, $A(r) \sim r^{-\beta_A}$ $L(r) \sim r^{-\beta_L}$ and $V(r) \sim r^{-\beta_V}$.

However what is of particular interest is the way these geometric measures relate to one another as their overall sizes change. This is allometry. The critical hypothesis is that as the size of the typical elements change, these relations may well depart from the standard geometric relations that characterize length, area, and volume. The allometric hypothesis suggests that there are critical ratios between geometric attributes that are fixed by the functioning of the element in question and if the element changes in size, these ratios need to remain fixed for the element to still function. Often the geometry has to change if these ratios are fixed [24]. A good example relates to natural light penetrating buildings. As natural light depends on the surface area, then to preserve a given ratio of natural light for the volume of the building, the shape of the building has to change if the building is increased in size. In short, the surface area does not change at the same rate as volume and if the ratio has to be fixed to make the building function, then the volume has to change. This implies a change in shape as the building increases in size.

As yet there is no well worked out theory of urban allometry; indeed there is no complete theory of size in biological systems from whence these ideas arise [25] although there are various theories in the making [14]. We will begin by stating basic geometric relations now assuming an idealized building to be a cube with its basic linear unit as $L$. $L$ first determines the area $A$ as $L^2$ and then volume $V$ as $L^3$ from which it is clear that $V = A\,L$. Standard allometric relations first proposed by Huxley [26] can be immediately derived which imply changes in the volume, area or length relative to each other of these measures. For our cube (which can be easily generalized to a less uniform geometry), $A = V^{2/3}$, $L = A^{1/2}$, and $L = V^{1/3}$. These imply that as the volume grows, the area grows at a rate 2/3rd's the rate of volume growth. This can easily be seen in the relative growth rate or ratio of $dA/A$ to $dV/V$ (assuming a unit of time) as follows

$$\frac{dA}{dV} = \frac{2}{3} V^{(2/3)-1} = \frac{2}{3}\frac{V^{2/3}}{V} = \frac{2}{3}\frac{A}{V} \qquad . \tag{6}$$

Rearranging terms in (6), we get the ratio, the relative growth of $dA/A$ to $dV/V$, as

$$\frac{dA}{A} \bigg/ \frac{dV}{V} = \frac{2}{3} \qquad , \tag{7}$$

which can be easily generalized for any scaling parameter $\lambda$. The general allometric relation relating some physical property $y$ of an object to another $x$ is thus

$$y = G x^{\lambda} \qquad , \tag{8}$$

where the scaling parameter is the relative growth rate of $y$ to $x$

$$\lambda = \frac{dy}{y} \bigg/ \frac{dx}{x} \qquad . \tag{9}$$

$\lambda$ is also the elasticity as defined in economics. Equations (8) and (9) can thus be applied to any relationship which might be scaling with respect to different measures of size where these sizes imply differential relative growth [27].

To simplify our treatment, we assume that the entire array of buildings can be represented as blocks based on polygonal footprints with a standard height. In fact this is the case as we will see in our buildings data base where buildings are constructed from plot area and mean height and where more complex buildings are glued together from simpler blocks. Then in terms of building blocks, linear dimension will involve heights $H_j$ in the (**z**) dimension and vector lengths in the (**x, y**) plane from which the area of the plot $A_j$, its perimeter $L_j$, and its volume (or mass) $V_j$ can be computed. We will not compute surface area of the building, or any internal measures of circulation or areas of interior space. These are not yet possible although currently the databases are being augmented to deal with such complexities. These four measures are defined for each building which is located at a point or centroid $j$ (or toid in the jargon of the relevant geography[1]). We are interested in their scaling with respect to rank-size which we have hypothesized above but we are also interested in how they scale with respect to each other, allometrically. The following scaling relations are stated:

$$\left.\begin{array}{r} H_j = Z_1 L_j^{\kappa}; \quad A_j = Z_2 L_j^{\eta}; \quad V_j = Z_3 L_j^{\mu} \\ A_j = Z_4 H_j^{\varphi}; \; V_j = Z_5 H_j^{\chi} \\ V_j = Z_6 A_j^{\vartheta} \end{array}\right\} \quad . \qquad (10)$$

where the $Z_*$ are the constants of proportionality and the power symbols are the appropriate allometric parameters – relative rates of change.

[1]A Toid (TOpographic IDentifier) is a unique reference identifier for every map feature in the UK, see http://www.ordnancesurvey.co.uk/oswebsite/freefun/geofacts/geo1201.html.

Our key interest in urban allometry is to find out whether the scaling between area and volume implies changes in the shape of buildings. In terms of the relations in (10), we would expect the volume to scale as the cube of height and perimeter, and as the square of the plot area. Plot area is likely to scale as the square of height and perimeter while perimeter and height scale with each other linearly. These are the baseline allometries that we might expect. However if there are changes of shape, then these will reflected in the parameter values that are estimated from the equations in (10). In fact, as it is likely that there will be considerable variation around these forms for all buildings, we will disaggregate the set of all buildings into different land use types which should reveal differences, particularly between buildings in commercial and residential use.

Currently we are not able to measure the surface area of building from the database and this is unfortunate as this may scale quite differently from the 2/3rd's ratio that pertains to the standard allometric equations. This is because the skin of the building is the conduit for light and energy. Buildings cannot maintain their volume indefinitely through increasing their floor areas because such areas cannot be serviced through natural light and other forms of externally supplied energy. Thus there are limits on shape in this regard. This is why it is likely that as buildings increase in size, they expand vertically rather than horizontally which are the kind of deviations from standard allometry that we are seeking. Our ultimate concern in this work is to count the number of building types by land use and to link these counts and their shapes to energy emission in buildings as well as issues involving circulation both within and between buildings.

## 4 Building data and the preliminary analysis of heights

To show that scaling exists in the size of buildings, we begin by selecting height data for the top 200 buildings worldwide and compare these with the same number for London. These data are from the **Emporis** database (http://www.skyscraper.com/) which contains quite detailed information about the largest buildings in 50,000 cities worldwide with up to 3000 of the largest buildings from the largest cities. We will also examine three cities in more detail – London, New York and Tokyo – as a prelude to our work with the Greater London buildings database which we outline below. This is taken from our Virtual London model which consists of building blocks constructed from digital data sources. In Fig. 1(a), we show Zipf plots of the top 200 buildings by height worldwide, for London from the **Emporis** database, and for London from our own database. We have also graphed the top 200 cities by population in the year 2006 taken from UN sources to show that scaling in population is a little more extreme than for high buildings. In all the Zipf plots that we introduce henceforth, we normalize the data in the following way. We normalize the rank $r$ by dividing by its maximum $r_{\max}$ and for the size variable, height $H_j$ say, we divide by its mean $< H_j >$. Our plots are then based on graphing $H_j / < H_j >$ against $r / r_{\max}$, thus enabling us to directly compare data by collapsing all the plots onto one another.

a) b)

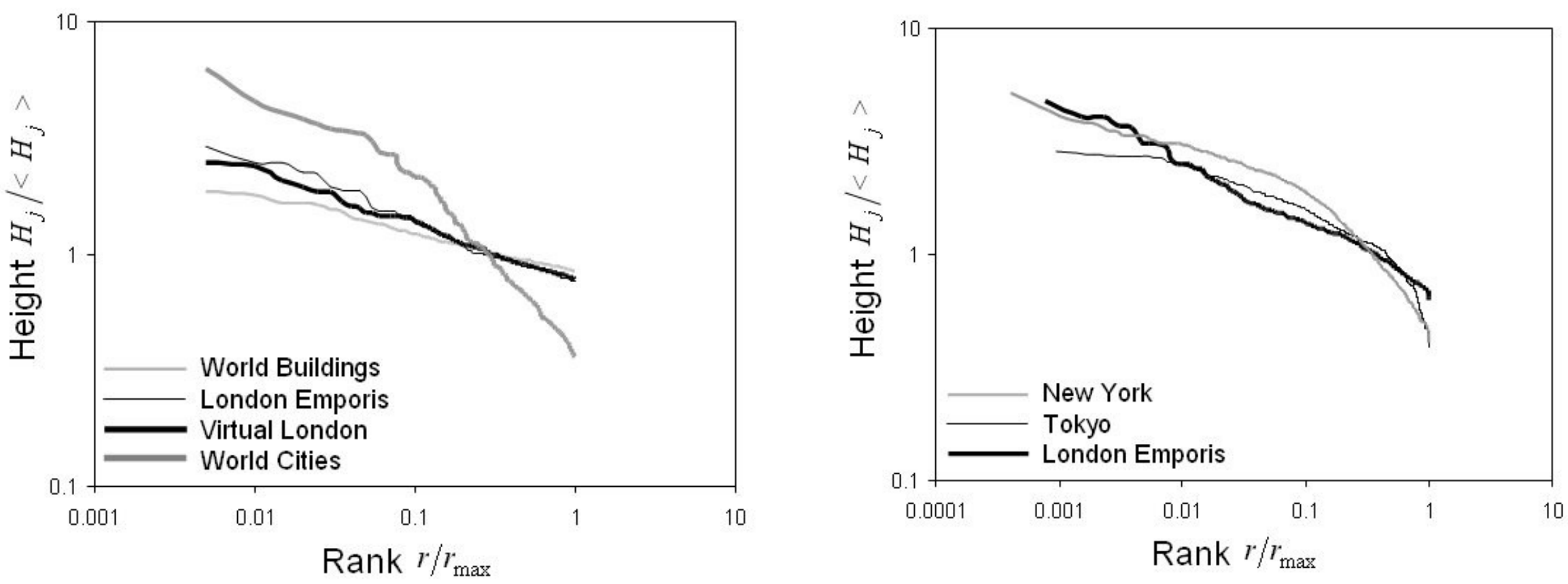

*Fig. 1. Initial analysis of building heights*

a) Top 200 buildings by height in the World and London, and top 200 city populations b) Top building heights in New York, Tokyo and London

There is very clear scaling in all four data sets and we present the parameters of these in Table 1. The slope of the world cities data is steeper than the buildings data which implies that there is less competition for activity inside these cities than between them. We have also examined the same scaling in building heights for three world cities from the **Emporis** database and in Fig. 1(b) we show these building heights over a wider range of magnitudes for Tokyo, London and New York. The results which are also shown in Table 1 imply that New York has greater competition than Tokyo and that London has the flattest profile in terms of rank-size scaling. Although the fit of the power law to the London and World data sets is good, this is less so for Tokyo and New York where there is clear evidence of lognormality in the plots even at their upper end. This simply confirms the observations made above about needing to exercise care in approximating such urban distributions by power laws.

| | *world cities(1)* | *world buildings (2)* | *London **Emporis** (2)* | *Virtual London (3)* | *Tokyo (2)* | *London (2)* | *New York (2)* |
|---|---|---|---|---|---|---|---|
| $N$ | 200 | 200 | 200 | 200 | 1036 | 1302 | 2424 |
| scaling parameter $\beta_H$ | 0.652 | 0.162 | 0.262 | 0.234 | 0.377 | 0.288 | 0.478 |
| correlation squared | 0.970 | 0.995 | 0.983 | 0.992 | 0.827 | 0.979 | 0.919 |
| density parameter $\alpha$ | 2.534 | 7.159 | 4.823 | 5.269 | 3.650 | 4.477 | 3.094 |

*Table 1. Scaling parameters for the preliminary analysis of building heights*

Sources: *(1)* United Nations (http://unstats.un.org/unsd/) *(2)* Emporis (http://www.skyscraper.com/), and *(3)* Infoterra (http://www.infoterra-global.com/)

This preliminary analysis gives us some confidence that there is scaling in building geometries leading us to develop the analysis of the much larger database for London based on our 3-D GIS/CAD model of London which we refer to as Virtual London [28]. This is a digital model of all building blocks within about 40 kilometers of the CBD – the City of London or 'square mile' – covering the 33 boroughs comprising the Greater London Authority (GLA) area which has an extent of 1579 square kilometers. The data set is unique in that it has been created automatically from two main sources of data: first vector parcel files from Ordnance Survey's **MasterMap** which code all land parcels and streets to about one meter accuracy (http://www.ordnancesurvey.co.uk/oswebsite/products/osmastermap/); and second a data set of buildings heights constructed from InfoTerra's LIDAR data which produces a massive cloud of 3-D **x-y-z** data points which when used in association with the vector parcel data, can be used to extrude all buildings. In this data set, there are some 3,595,689 ($\cong$ 3.6m) distinct building centroids (toids). We are currently dealing with all 3.6 million although we only use a subset of these in our scaling and allometric analysis. In future work, we will be aggregating toids to ensure that we are dealing with appropriate blocks. This becomes critical when land use is to be assigned to each building block because land use is tagged to street addresses which are a subset of all toids.

To give some idea of the range of this data set, the maximum height of any block is 204.06 meters, the Canary Wharf Tower in the London Docklands. The mean height is 5.76 meters and the standard deviation is 3.29 meters which shows that the frequency of building heights is very skewed to the left, reflecting the fact that this distribution is likely to follow a power law[2]. For illustrative purposes only the top 10 blocks by height in London are 204, 197, 169, 160, 151, 150, 138, 130, 128, and 123 meters in comparison with the top 10 from the **Emporis** world database which are 509, 452, 452, 442, 421, 415, 391, 384, and 381. London's highest building is not in fact in the top 200 in the world and from the regression in Table 1 associated with the plot in Fig. 1(a), we can estimate its rank as about 400. London is not a city of tall buildings.

From the data set, we are currently working with the perimeter of each plot which is computed directly from the **MasterMap** data, and the mean height of a plot which is important as there are many different heights from the LIDAR data reflecting complex roof shapes, masts, air conditioning units and so on. Other measures of height such as median and mode do not change the results below substantially. We compute volume by taking the area of the plot and multiplying it by its height. This does not take account of course of the fact that some buildings will taper but currently we are not able to do much about this as we do not have elaborate algorithms in place to construct intricate roofing shapes. We also are able to classify these buildings by land use from the **MasterMap** Layer 2 where we have land uses associated with each street address for which there is a toid. However there are many blocks that do not have street addresses and these tend to be part of other building complexes and/or are very small and somewhat idiosyncratic in their form, such as sheds, lean-to's and such-like bric-a-brac. We have various algorithms for joining unclassified polygons to those which are already classified and currently we consider the data set to be robust.

[2] This relatively low average height compared to the largest building in Greater London simply illustrates that the database is dominated by low rise residential properties.

There are over six hundred different land use types in the **MasterMap** data and we have classified these into nine major types which we list as the toids classified with at least one residential, office, retail, services, industrial, educational, hotel, transport, and general-commercial land use. We have not yet broached the difficult question of multiple uses for if we have a building with more than one land use classifier, we simply include it in the appropriate analysis. We have not yet tackled this double counting.

## 5 Rank-size distributions and allometric analysis of building geometries

We begin with the aggregate scaling relations which result from ranking the area $\{A_j\}$, perimeter $\{L_j\}$, height $\{H_j\}$ and volume $\{V_j\}$ data for a slightly reduced data set of about 3.58 million buildings. We show these in Fig. 2 which also contains the same scaling for each of the land uses which we will describe below. What this figure reveals is remarkably strong linearity over many orders of magnitude with the plots collapsing dramatically for the million or so smallest buildings (which are less than about 25 square meters in volume) and quite definitely represent the bric-a-brac of urban construction picked up from the remote sensing. These plots do not show any lognormality which is perhaps surprising given other size distributions [20, 23] and when the right tail is excluded from the data, the linearity is even more apparent. In fact what we have done in fitting power laws to these data is fit the generic equation to only the top 10 percent of buildings.

The aggregate plots are shown in the thick black line in Figs. 2(a) to (d) with the excluded data points in grey. We have estimated the scaling parameters $\beta_A$, $\beta_L$, $\beta_H$, and $\beta_V$ from the appropriate rank-size equations using log-linear regression but we must note that as volume is a simple product of area and height $V_j = A_j H_j$, then this is a derived variable that does not have the same status as the raw data variables area, perimeter and height. In fact area and perimeter are confounded too as perimeter and area are both formed from the same two linear dimensions defining the

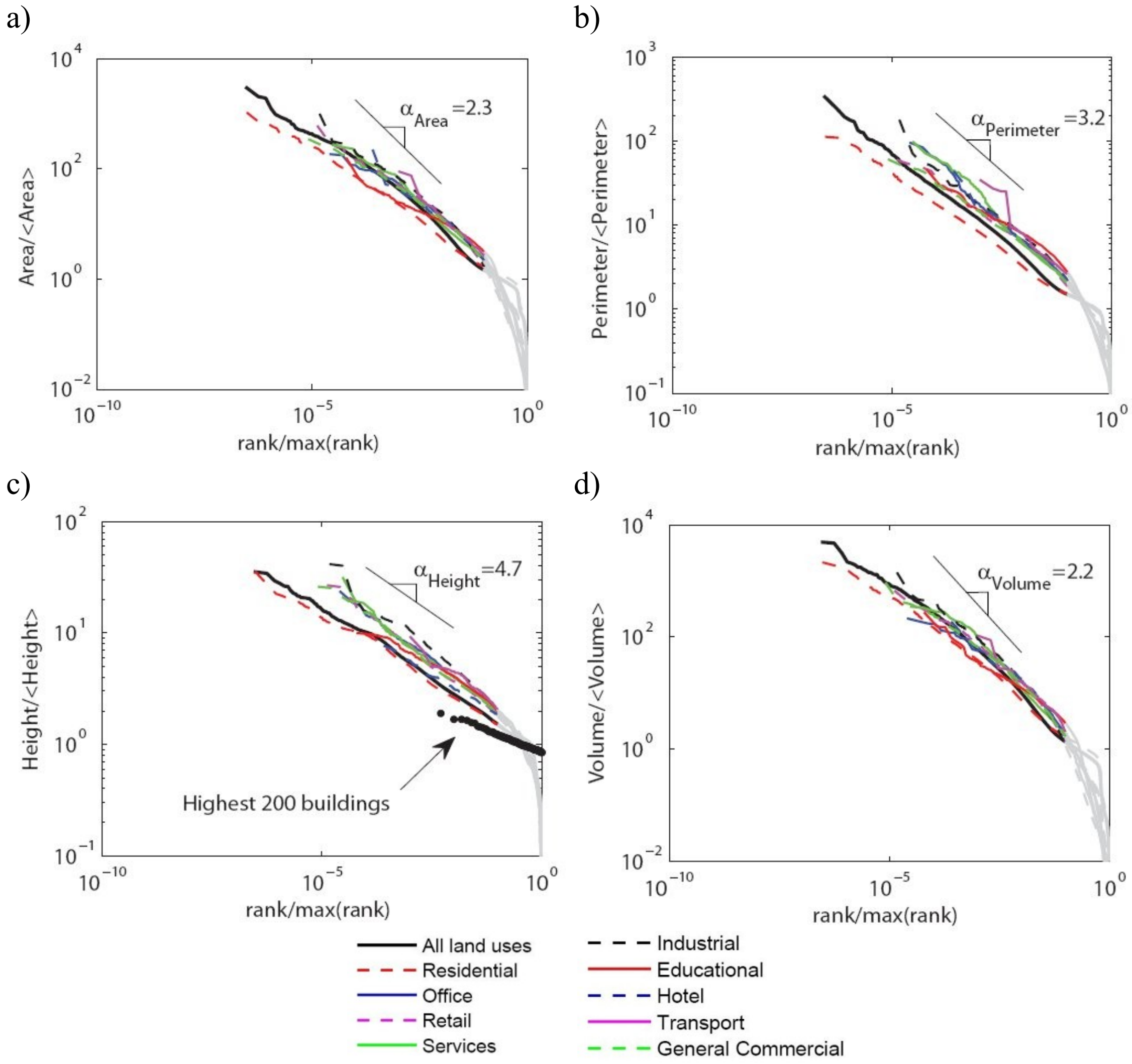

*Fig. 2. Normalized rank-order plots*
*a) building area, b) perimeter, c) height and d) volume*

We plot all buildings (solid curves in black) and buildings classified by their land use (dashed and dotted curves). We also plot fits to the rank-size distribution for all buildings (all land uses) on each panel and compute the corresponding regression coefficients applying the least squares method to the top 10% ranks in each curve. Panel c) includes the rank-order plot of the height for the highest 200 buildings from the **Emporis** worldwide database rectangular blocks that make up the buildings data set. We include volume and area because these are two variables that are usually used in describing cities, notwithstanding the fact that they are composed of more basic geometric primitives. To illustrate the interdependence between these results, if the rank order $r$, for height and for area, were identical, that is for $A(r) \sim r^{-\beta_A}$ and $H(r) \sim r^{-\beta_H}$, then volume could be predicted as $V(r) \sim r^{-\beta_A} r^{-\beta_H}$. This is unlikely to be the case for we know that height is likely to increase faster than area as buildings seek space upwards. In short, this is why we need to examine the allometric relations which relate the various quantities. Thus we might expect volume to decline more steeply with rank than area, which in turn is likely to fall more steeply than height or perimeter for this is the sequence of objects from 3 to 2 to 1 dimension.

In Table 2, we present the results which also show the data for same scaling relations for the land uses. We have very dramatic linearity in the log-log plots over several orders of magnitude for volume from $10^7$ to $10^2$ after which the plot falls very steeply, implying that buildings less than 25 square meters in volume behave quite differently. These are really sheds and bric-a-brac referred to earlier and in future work will be discounted to an extent as we construct better building blocks [29]. These regressions are striking in their linearity and such rank-size relations are amongst the best we have come across. In fact this bears out the remarkable linearity of the rank-size of the heights of the top 200 buildings in the world which enabled us to make such good predictions of building heights further down the scale. The rank-size plots for the nine land use categories – residential, office, retail, services, industrial, educational, hotel, transport, and general-commercial – are also shown in Fig. 2 with respect to area, perimeter, height and volume. We expected these plots to show rather different scaling from the aggregate (although 90 percent of the buildings in the database are classified as residential land use) but they are all close to the aggregate relations. From Fig. 2, it is clear that their linearity tends to be over a lesser number of orders of magnitude. Any differences that do occur in these slopes are highlighted in Fig. 3 which compares the $\beta$ coefficients and their error bars.

| | *all land uses* | *resid-ential* | *office* | *retail* | *services* | *indus-trial* | *educa-tional* | *comm-ercial* |
|---|---|---|---|---|---|---|---|---|
| $N$ | 3595689 | 3320579 | 39587 | 77075 | 33949 | 67270 | 16257 | 122874 |
| area $\beta_A$ | 0.763 | 0.559 | 0.711 | 0.802 | 0.664 | 0.840 | 0.486 | 0.711 |
| perimeter $\beta_L$ | 0.272 | 0.251 | 0.294 | 0.305 | 0.308 | 0.352 | 0.272 | 0.287 |
| height $\beta_H$ | 0.457 | 0.352 | 0.457 | 0.461 | 0.469 | 0.477 | 0.393 | 0.432 |
| volume $\beta_V$ | 0.861 | 0.688 | 0.834 | 0.923 | 0.841 | 1.007 | 0.570 | 0.841 |

*Table 2: Scaling Parameters for Buildings in the London Database*

Note that only the top 10 percent of these building numbers are used in the regressions and that Transport and Hotel have been excluded due to their small numbers

The six sets of allometric relations stated earlier in equations (10) are plotted in logarithmic form as two-dimensional surfaces in Fig. 4. Only three of these relationships show the kind of linearity that we might expect from our earlier analysis, and these involve area v. perimeter, volume v. perimeter and volume v. area, that is those based on $A_j = Z_2 L_j^{\eta}$, $V_j = Z_3 L_j^{\mu}$, and $V_j = Z_6 A_j^{\vartheta}$. The other relationships involving height are quite scattered and require different techniques for extracting their allometry for clearly the set of data points must be culled to extract those that reflect the densest parts. As there are almost 3.6 million points in this scatter, their representation as surfaces colored by their density after appropriate binning into a relatively fine scale set of categories is the most useful way of assessing these relationships. In Table 3, we present results from estimating the three allometric regression lines to the data in its logarithmic form.

.

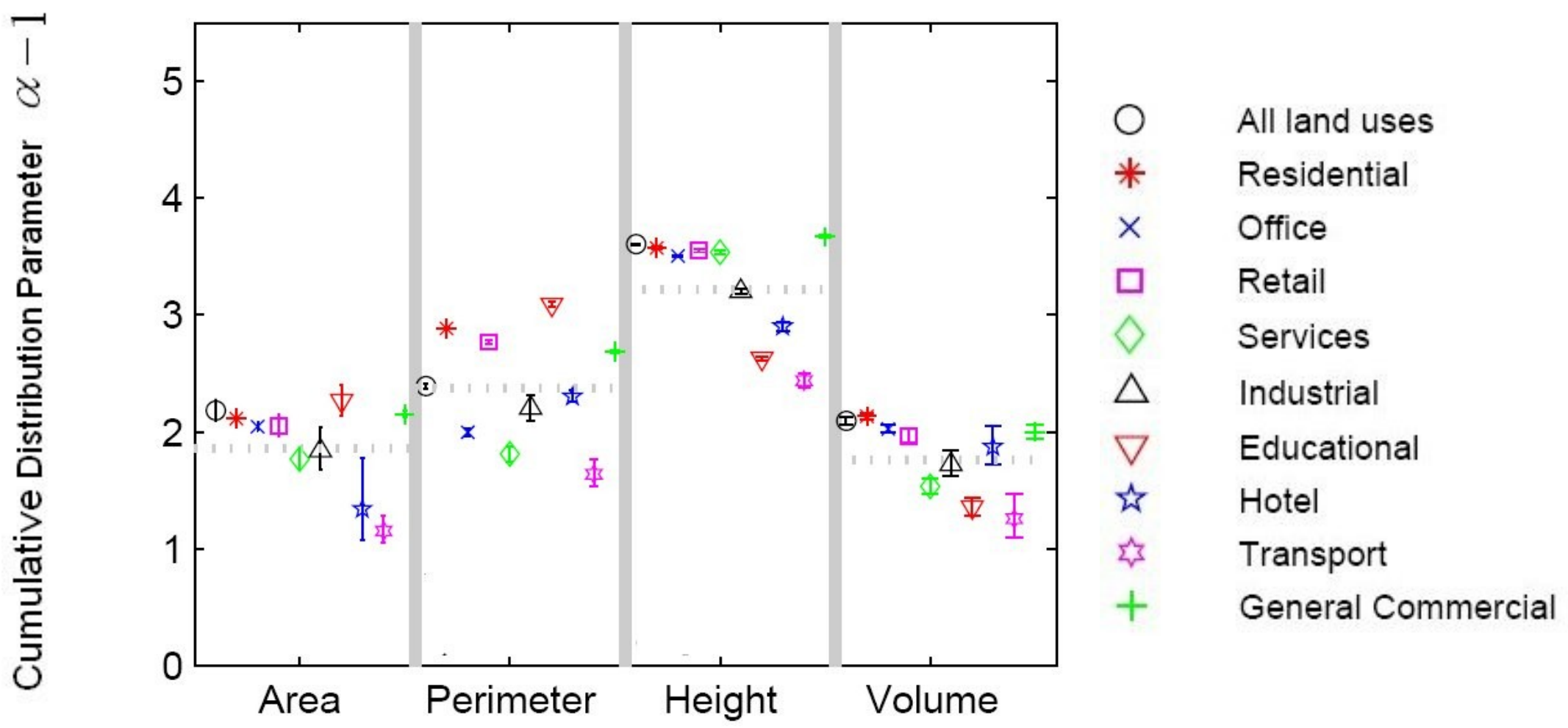

*Fig. 3. Scaling coefficients for the plots in Fig. 2.*

The least squares method is applied to the top 10% ranks in each curve. Error bars are 95% confidence intervals. Horizontal dashed lines (in grey) are mean values of $\alpha - 1$.

It is immediately clear that the values of these parameters are consistent with the order of their geometric scaling. That is, the parameter of area on perimeter is less the square while the value of the relation between volume and area is less than 3/2. This means that as the perimeter increases, the area increases less than the normal geometric relation implying that shape is changing, probably becoming more crennelated – implying a longer perimeter – as the area grows. In terms of volume, this increases at less than 3/2 of the area which suggests that the volume must get proportionately less as the area grows. This bears out the implied observation that as the surface grows, the shape must change.

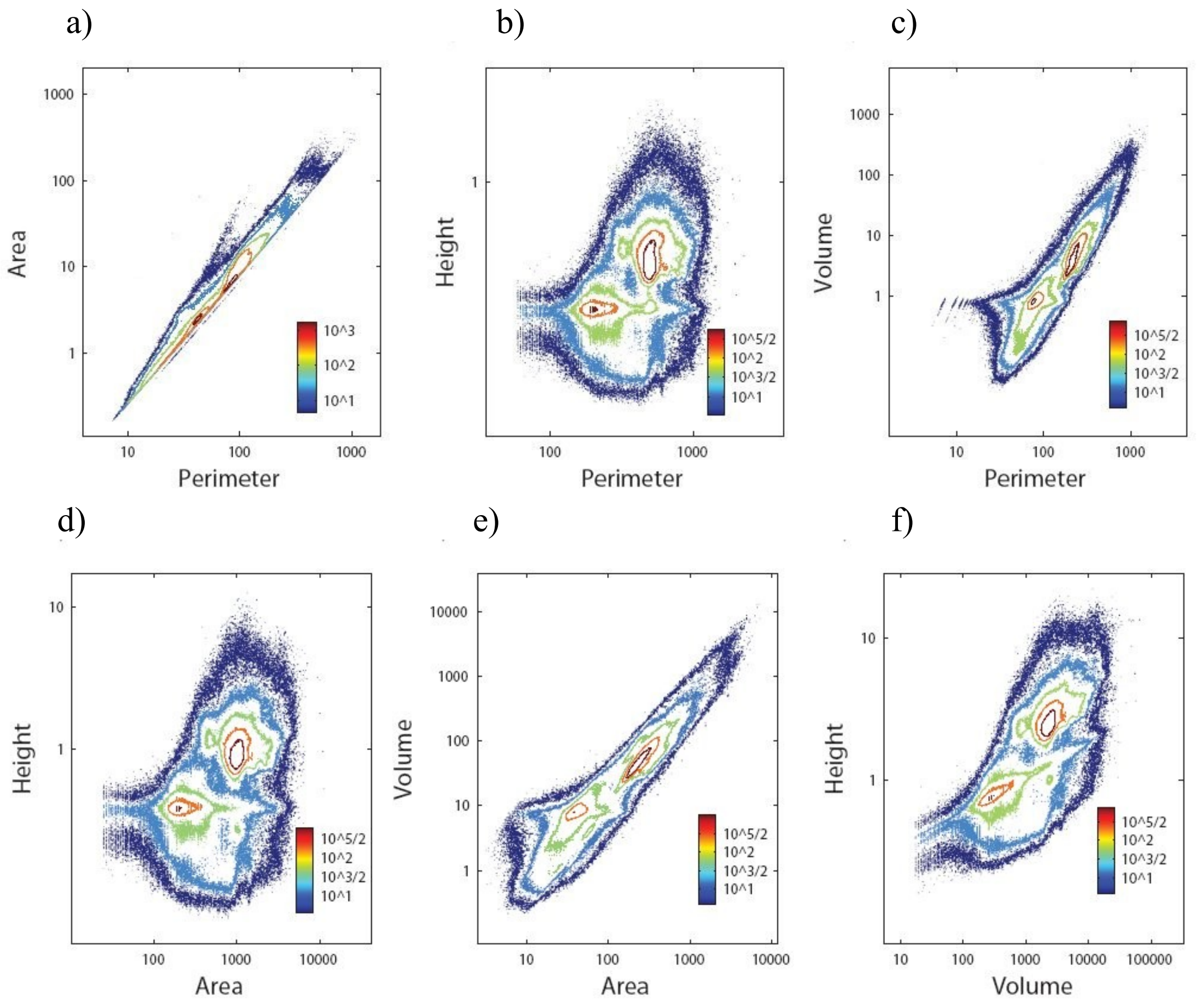

*Fig. 4. Two-dimensional surface plots of allometric relations*

a) Perimeter against Area, b) Perimeter against Height, c) Perimeter against Volume, d) Area against Height, e) Area against Volume and f) Volume against Height. Each panel implies contour lines of logarithmically-binned histograms (frequency counts) on a logarithmic scale. Color bars display the range of histogram values on each panel. We have found an approximate linear relation between the variables in panels a), c) and e).

Table 3 also contains all the parameter estimates for these three relationships for each individual land use. Remarkably these are <u>all</u> consistent with the aggregate and show that building volumes grow proportionately less than their increases in area as we might expect. However, it is even more urgent now to extend the analysis of allometry to height as this is a key variable in defining volume and it is the weakest aspect of our work, largely because we cannot assume that usable building volumes are the same as geometric volumes. Moreover the height data itself is highly variable due to the fact that we have used mean height which is not necessarily a good measure for computing volume. This requires considerable further research as it is central to some of the notions in this paper which relate to how volume scales with plot area and to questions of surface area that define building skins. This is the research we will develop next when we link the buildings database to related databases of floorspace and energy emissions.

| | **area v perimeter** | | **volume v perimeter** | | **volume v area** | |
|---|---|---|---|---|---|---|
| | allometric coefficient | r-square | allometric coefficient | r-square | allometric coefficient | r-square |
| Euclidean scaling | 2 | | 3 | | 3/2 | |
| all land uses | 1.832 (1.832, 1.833) | 0.962 | 2.386 (2.385, 2.387) | 0.811 | 1.296 (1.296, 1.297) | 0.835 |
| residential | 1.846 (1.846, 1.846) | 0.963 | 2.463 (2.461, 2.464) | 0.825 | 1.325 (1.324, 1.326) | 0.845 |
| office | 1.783 (1.779, 1.787) | 0.952 | 2.152 (2.141, 2.162) | 0.808 | 1.199 (1.194, 1.204) | 0.838 |
| retail | 1.811 (1.808, 1.814) | 0.958 | 2.215 (2.207, 2.222) | 0.816 | 1.216 (1.212, 1.219) | 0.842 |
| services | 1.773 (1.769, 1.777) | 0.964 | 2.129 (2.118, 2.140) | 0.814 | 1.195 (1.189, 1.200) | 0.836 |
| industrial | 1.788 (1.786, 1.792) | 0.957 | 2.052 (2.042, 2.062) | 0.706 | 1.148 (1.142, 1.153) | 0.738 |
| educational | 1.679 (1.673, 1.684) | 0.959 | 1.901 (1.888, 1.914) | 0.828 | 1.132 (1.125, 1.139) | 0.862 |
| hotel | 1.770 (1.760, 1.780) | 0.969 | 2.143 (2.115, 2.172) | 0.849 | 1.207 (1.193, 1.222) | 0.870 |
| transport | 1.775 (1.749, 1.801) | 0.948 | 1.991 (1.928, 2.053) | 0.797 | 1.116 (1.085, 1.147) | 0.833 |
| general commercial | 1.813 (1.811, 1.815) | 0.956 | 2.179 (2.173, 2.185) | 0.813 | 1.194 (1.191, 1.197) | 0.840 |

*Table 3. Coefficients and correlations for the allometric relations*

The numbers in brackets in the coefficient columns give the 95% confidence intervals

## 6 The spatial distribution of building geometries

To put space back into the argument, we can examine the two-dimensional distribution of building geometries in Greater London by computing the correlation functions with respect to how properties of a building – area, height and so on – vary with respect to every other building. From our previous analysis of the spatial distribution of population densities with respect to how density varies a centre point which invariably declines exponentially with distance from the CBD [4], we might except that these correlation functions to imply power laws with respect to increasing distance from any building in question. In this section, we will compute a composite correlation function in the following way, assuming that building properties meet the definitions of a point process.

The first moment of such a point process can be specified by a single number, the intensity $\rho$ giving the expected number of points per unit area. The second moment can be specified by Ripley's $K$ function [30] where $\rho K(R)$ is the expected number of points within distance $R$ of an arbitrary point of the pattern. The product density

$$\rho_2(x,y)dA(x)dA(y) = \rho^2 g(R)dA(x)dA(y) \qquad (11)$$

describes the probability of finding a point in the area element $dA(x)$ and another point in $dA(y)$, at distance $R = |x-y|$, and $g(R)$ is the two-point correlation function. Ripley's $K$ function is related to $g(R)$ as

$$K(R) = 2\pi \int g(R)dR \quad . \qquad (12)$$

In other words, $g(R)$ is the density of $K(R)$ with respect to the radial measure $RdR$ [31]. The benchmark of complete randomness is the spatial Poisson process, for which $g(R)=1$ and $K(R)=\pi R_2$, the area of the search region for the points. Values larger than this indicate clustering on that distance scale, and smaller values indicate regularity.

The two-point correlation function can be estimated from $N$ data points $x \in D$ inside a sample window $W$ as

$$g(R) = \frac{|W|}{N(N-1)} \sum_{x \in D} \sum_{y \in D} \frac{\Phi_R(x,y)}{2\pi R \Delta} \omega(x,y) \quad , \qquad (13)$$

where $2\pi R\Delta$ is the area of the annulus centred at $x$ with radius $R$ and thickness $\Delta$ [32]. Here $|W|$ is the area of the sample window, and the sum is restricted to pairs of different points $x \neq y$. The function $\Phi_R$ is symmetric in its argument and $\Phi_R(x, y) = [R \leq d(x, y) \leq R + \Delta]$ where $d(x, y)$ is the Euclidean distance between the two points and the condition in brackets equals 1 when true and 0 otherwise. The function $\omega(x, y)$ accounts for a bounded $W$ by weighting points where the annulus intersects the edges of $W$. There are a number of edge-corrections available, but that developed by Ripley [33] has a long tradition both in human geography and physics [34]. Here we approximate $\omega(x, y) = 1$ as the city does not have clear spatial boundaries.

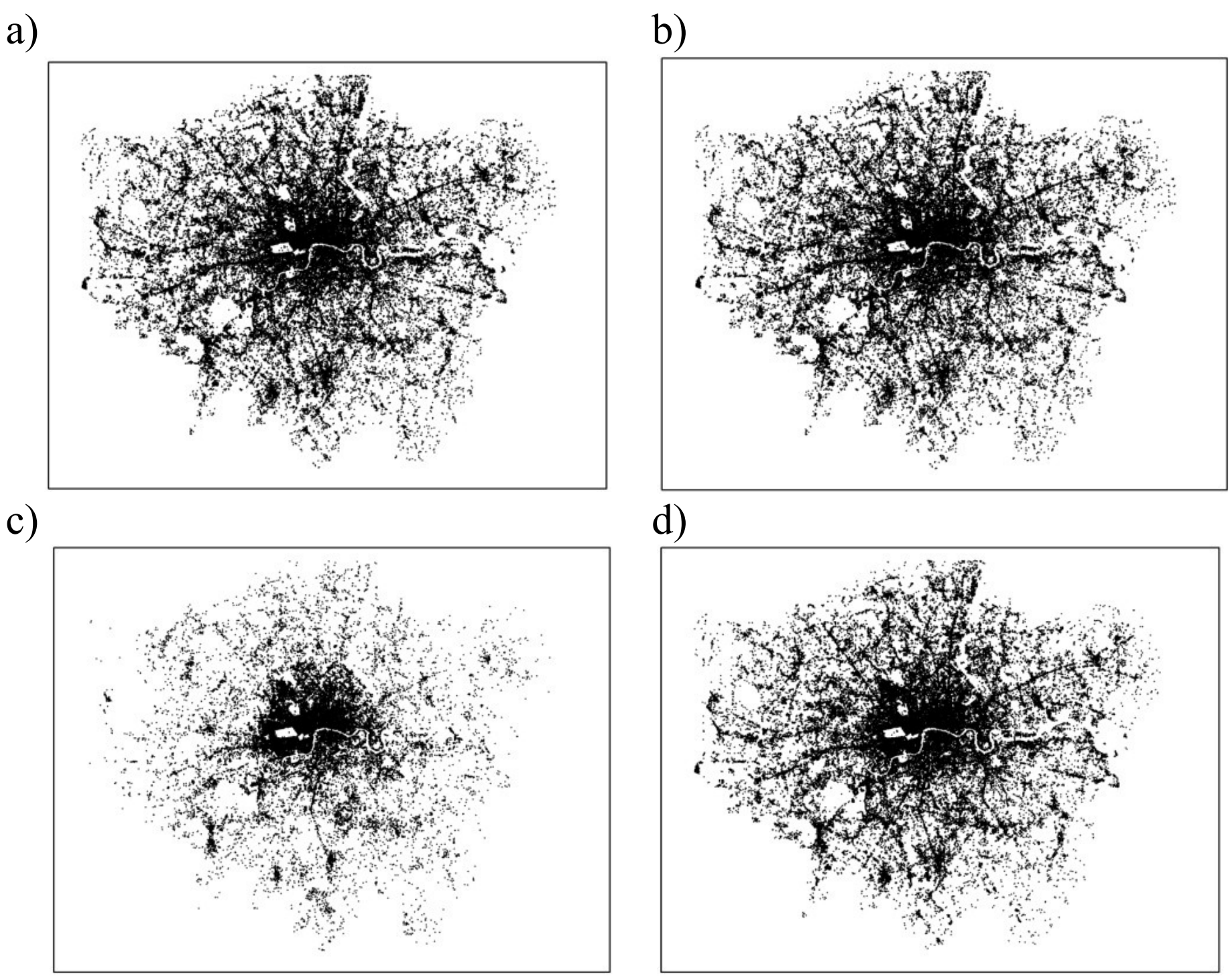

*Fig. 5. Spatial distribution of the geometric properties of the highest 100,000 buildings*

a) area b) perimeter c) height, and d) volume

Of special physical interest is whether the two-point correlation is scale-invariant. A scale-invariant $g(R)$ is an indicator of a fractal distribution of points, and is expected in critical phenomena [32]. Fig. 5 shows the distribution of geometric building properties over Greater London for the largest 100,000 selected by height for a range of distances up to $r \cong 3.3\,\text{km}$ and Fig. 6 is a plot of the two-point correlation function on a double logarithmic scale. We observe a power-law decay of $g(R) \sim r^{-0.230} \sim r^{-\gamma}$ for these largest 100,000 buildings. Interestingly, the two-point correlation function does not display scaling behaviour if we select the 100,000 largest buildings by perimeter size or area. This suggests that building height is a major variable which has so far been overlooked in studies of the fractality of cities and this supports our preliminary analysis of height from related databases.

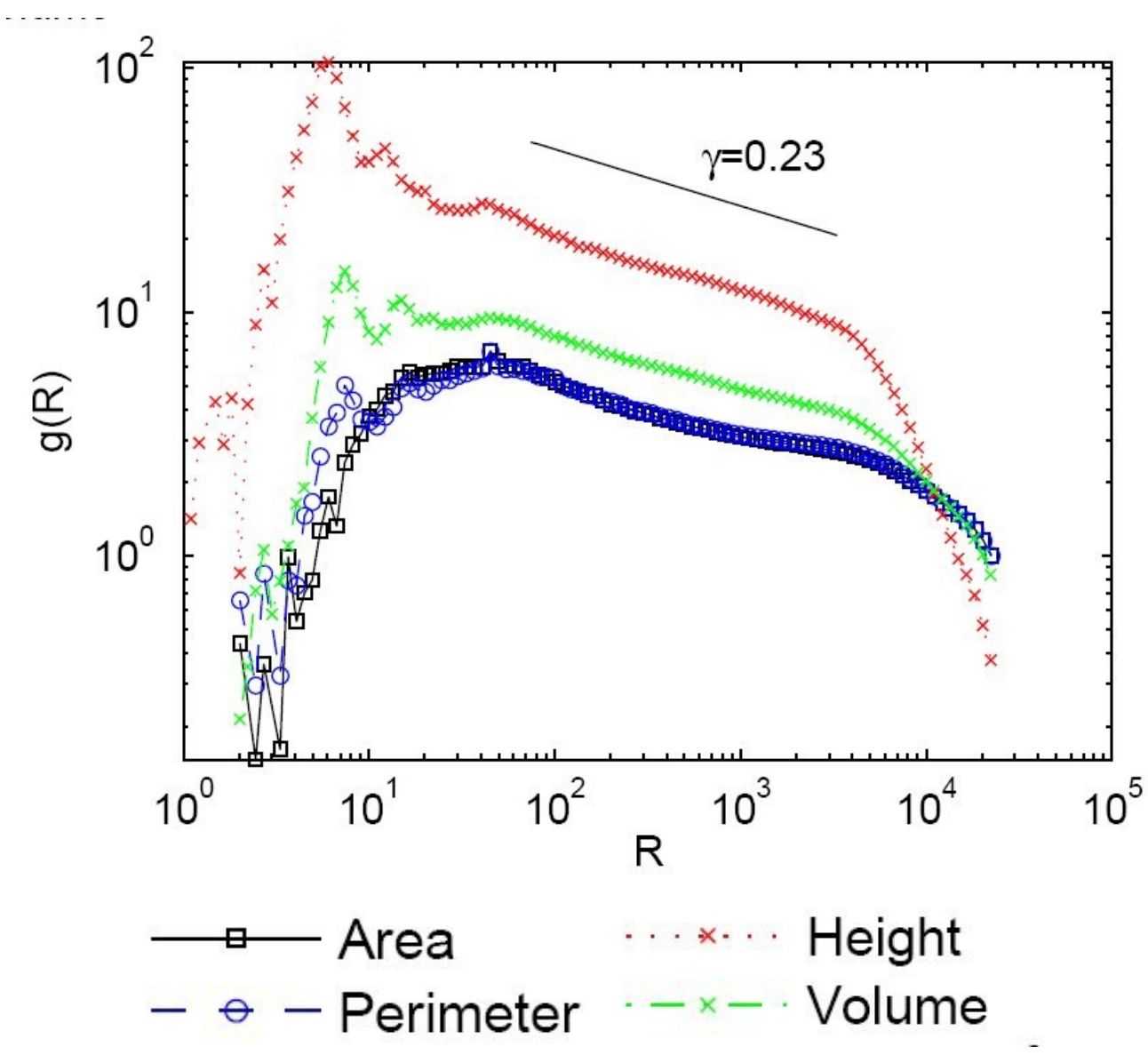

*Fig. 6. Two-point correlation functions of the building geometries with respect to distance R*

.

## 7 Next Steps

Our analysis represents a first step in developing scaling and allometry for spatial distributions within cities and this suggests a research program complementary to that being developed for equivalent relationships between cities [15]. The link between the rank-size scaling of spatial attributes which suppresses the spatial pattern and the scaling of the spatial patterns with respect to distance which we briefly introduced in terms of two-point correlation functions, needs to be explored in considerably more depth. We also need to investigate the relationship between geometric and socio-economic attributes as reflected in the link between building geometries and population densities as this serves to link the physical form of the city to its functioning. Definitional problems abound when data which is spatial are explored. Data based on individual objects such as people frequently does not display spatial pattern until it is aggregated. Although attributes such as income do accord to scaling at the level of individuals, many others are only retrieved when the data is aggregated to some specific level and thus the degree to which it is aggregated is critical. We need to revisit these definitional issues in more detail and in the case the database used here, iron out many of the problems of building size and type that we have identified. The analysis should be extended to deal with different rank-size and allometric relations in different areas of the city, showing how these relations might change as implied in the distributions pictured, for example, in Fig. 6.

We are much encouraged by the very strong scaling implicit in this data. Of course to confirm this, we need more examples from other cities. We need to relate the physical geometry to other measures, particularly linear measures such as utilities and street systems as well as socio-economic activity volumes as proposed by Kuhnert, Helbing, and West [9] amongst others. We need to link the analysis much more strongly to fractal geometry [35] and we need to link it to circulation patterns in buildings [6-7]. We will examine the surface areas of buildings linking these to energy emissions and related phenomena and when we do this, the variations in these relations with respect to different locations and districts within the city will take on new meaning. In time, we hope that such work will add to our growing knowledge of how efficient cities are in terms of their geometry and in this sense, provide a much more considered position on issues such as urban sprawl and the compact city.

## 8 Summary

In summary, we define and fit power laws and allometric scaling relations to four geometrical properties of buildings – perimeter and area of each building, plot, height and volume – for a large database of buildings in Greater London. We begin by defining how power laws approximate the underlying distributions which arise from competition for sites, and then we examine heights for the top 200 buildings world wide and for buildings in three world cities, New York, Tokyo and London. We then develop this analysis for the London data and demonstrate strong scaling in terms of rank-size and significant scaling distortions with respect to allometric relations between area, perimeter height and volume. We conclude with suggesting that once we reintroduce space into these distributions using two-point correlations that the height distribution scales spatially with distance. This sets the agenda for further research.